\documentclass[12pt,preprint]{aastex61}
\usepackage{color}
\usepackage{CJK}
\usepackage{longtable}
\usepackage{graphicx}
\usepackage{epstopdf}

\newcommand{\msun}{{M}_{\sun}}

\shorttitle{The physical mechanism for CL AGNs} \shortauthors{Liu et al.}

\begin{document}

\title{Evidence for changing-look AGNs is caused by change of accretion mode}

\correspondingauthor{Qingwen Wu}
\email{qwwu@hust.edu.cn}

\author{Hao Liu}
\affiliation{School of Physics, Huazhong University of Science and Technology, Wuhan 430074, China}

\author{Qingwen Wu}
\affiliation{School of Physics, Huazhong University of Science and Technology, Wuhan 430074, China}

\author{Bing Lyu}
\affiliation{School of Physics, Huazhong University of Science and Technology, Wuhan 430074, China}

\author{Zhen Yan}
\affiliation{Shanghai Observatory, Chinese Academy of Science, Shanghai 200030, China}


\begin{abstract}
The discovery of changing-look active galactic nuclei (CL AGNs), with appearance and disappearance of broad emission lines and/or with strong variation of line-of-sight column density within a few years, challenges the AGN unification model. We explore the  physical mechanisms based on the X-ray spectral evolution for a sample of 15 CL AGNs. We find that the X-ray photon index, $\Gamma$, and Eddington-scaled X-ray luminosity, $L_{\rm 2-10 keV}/L_{\rm Edd}$, follow negative and positive correlations when $L_{\rm 2-10 keV}/L_{\rm Edd}$ is lower and higher than a critical value of $\sim 10^{-3}$. This different X-ray spectral evolution is roughly consistent with the prediction of the accretion-mode transition (e.g., clumpy cold gas or cold disk to advection dominated accretion flow, or vice visa). With quasi-simultaneous X-ray and optical spectrum observations within one year, we find that the CL AGNs observed with and without broad emission lines stay in the positive and negative part of the $\Gamma-L_{\rm 2-10 keV}/L_{\rm Edd}$ correlation respectively. Our result suggest that the change of the accretion mode may be the physical reason for the CL AGNs.
\end{abstract}


\section{Introduction}

Active galactic nuclei (AGNs) are believed to be powered by accretion of matter into a supermassive black hole (SMBH) that located at center of almost all massive galaxies. The AGNs are normally classified into Type I and Type II according to the width of the emission lines, where the Type I sources show both broad and narrow emission lines (BELs and NELs) while Type II sources only have narrow lines. The AGN unification models based on the orientation angle are widely accepted in  last two decades, where the type I AGNs are viewed face-on so that the BELs near the central SMBHs can be observed directly, while the BELs  in Type II AGNs are obscured by the putative parsec-scale torus if we are viewed edge-on \citep[e.g., ][]{up95}.  Once the optical type classified, the AGN is normally not expected to change between two types in a human timescale. However, several changing-look AGNs (CL AGNs) were discovered in last two decades.  It should be noted that the term `changing look' was originally referred to X-ray observations of Compton-thick AGN becoming Compton-thin, and vice-versa \citep[e.g., ][and references therein]{ricc16}. In optical spectrum, some AGNs undergo dramatic variability of the continuum, emission line profiles, classification types (e.g., move from one spectral class to another within a couple of years), which is also called as CL AGNs. With more and more multi-band surveys, the number of CL AGN is growing rapidly in last few years \citep[e.g., ][]{to76,mcel16,ya18,guo19,tr19}. 
 
 The physical mechanism for the strong variation of the line-of-sight column density from X-ray observations and type transition in optical spectrum is quite unclear, which is important for understanding the basic nature of AGNs. One scenario is that disappearing or emerging of BELs or the strong variation of the column density are ascribed to the variation of the clumpy torus, where the variable absorber moving in and out of the line of sight \citep[e.g., ][]{go89}. The second scenario is that the change of column density and the AGN classification type are caused by the intrinsic changes of AGN central engines.  The obscuration scenario should work in some CL AGNs based on the variation of the column density. However, some other CL AGNs cannot be explained by the obscuration scenario since that its intrinsic absorption is always low \citep[e.g., Mrk 1018,][]{mcel16}  and the low level of polarization in some CL AGNs  \citep[e.g.,][]{huts19}. Furthermore, the strong variation of the infrared emission and the time-lag between the infrared and optical emission also suggest that dust reprocessing of central optical/X-ray emission \citep[e.g.,][]{sh17}.  \citet{nd18} proposed that the CL AGNs are possibly triggered by the rapid drop or increase of accretion rate based on the broadband spectrum of a CL AGN (Mrk 1018) and it may be similar to the state transition in stellar-mass X-ray binaries (XRBs).  To explain the `time-scale problem' in CL AGNs,  \citet{sc19} proposed that the CL AGNs may be triggered by the radiation-pressure instability that occurred in a relatively narrow transition zone between the standard disk and optically thin advection dominated accretion flow (ADAF).

It is unclear that it is the same or not for the CL AGNs as defined from variations of column density and optical spectrum. Understanding the physics behind CL AGNs will play an important role in understanding the AGN physics and the life cycle of galaxies. In this work, we explore the physical mechanism of CL AGNs based on X-ray spectral evolution for a sample of CL AGNs, since that the X-ray emission is believed to come from the compact region near the black hole horizon and can shed light on the change of accretion or state. In particular, we also searched the optical spectrum information from the literatures, and define the quasi-simultaneous observation as that the X-ray and optical spectrum observations within one year.  We will learn the information about the change of broad emission lines with the change of accretion that derived from the X-ray observations. The sample selection and data analysis are described in Section 2. In Section 3 \& 4, we present and discuss the results. For this work, we adopt a cosmological model of $\Omega_{m}=0.32$, $\Omega_{\Lambda}=0.68$, and $H_{0}=67$ km s$^{-1}$ Mpc$^{-1}$ \citep[][]{plan14} in this work.

\section{Sample and data reduction}

We search the reported CL AGNs from literatures, where CL AGNs as defined from both variations of column density and change of the optical spectrum type are considered. To explore the possible evolution of the central engine of CL AGNs, we select the X-ray data for these CL AGNs from different X-ray telescopes. The archived data for both NuSTAR, XMM-Newton and Chandra are selected and reduced for four famous CL AGNs -- Mrk 1018, NGC 3516, Mrk 590 and NGC 2992, where these sources suffer strong X-ray variations in last 20 years.  We also reduced the NuSRAR archived data for another 8 CL AGNs. To investigate the statistical properties of CL AGNs, the X-ray photon indices of several other sources reduced from Suzaku, RXTE, Swift are also selected from the literatures. To understand the relation between the change of optical types of CL AGNs and the evolution of the central engine properties, we simply define the quasi-simultaneous data as that the X-ray and optical data are observed within one year.  The CL AGNs are classified as BEL CL AGNs, NEL CL AGNs and uncertain-type CL AGNs according to the quasi-simultaneous optical spectrum with BELs, with only NELs or uncertain type due to no quasi-simultaneous optical spectrum observations. The selected sources as well as their BH mass, redshift are listed in Table (1).

32 NuSTAR observations for 12 CL AGNs with the observation time longer than 10 kilo-seconds are selected. We reduce the data using the NuSTAR Data Analysis Software (nustardas v1.4.1), the CALDB version 20150316 and standard filtering criteria with the \textit{nupipeline} task.  The spectra are extracted by the \textit{nuproducts} task with a circular region of $60^{"}$ at the peak of the source and  $60^{"}$  radius circular background region away from the source. For spectral analysis, XSPEC (version 12.9.0) was adopted for the spectral fitting.  An absorbed power-law component (POW), and a reflection component (PEXRAV) are adopted, where the gaussian iron $\rm K_{\alpha}$ line at 6.4 keV is added if it is necessary.


The XMM-Newton observations are obtained from the instruments as EPIC PN and MOS camera. For these observations, we reduce the data and extract the spectrum using the Science Analysis Software (SAS, version 18.0.0), where the source and background region is same as that used in NuSTAR. 
30 observations for 5 sources observed by XMM-Newton from the archived data are selected. The data of both PN and MOS1/MOS2 are considered. In model fitting, the reflection component is neglected since it cannot be well constrained based on such narrow energy band.  One observation of NGC 2992 (ID0147920301) is thought as pile-up data, whose spectra is extracted from an annular source region with 10 and 40 arcseconds as inner and outer radius.


Three sources with Chandra observations are also explored, which are reduced by Chandra data analysis software (i.e., CIAO, version 4.11) with the CALDB (version 4.8.4.1) as calibration file.  The data from ACIS camera are adopted and \textit{chandra\underline{ }repro} and \textit{specextract} are used in our data reduction. Two observations(12868 and 4924) are considered as pile-up data, and an annular source region within 1 and 6 pixels as inner and outer radius are adopted in spectral analysis. Spectral fitting for these observations are similar to that of XMM-Newton as described above. The fitting parameters are listed in Table (2).

\section{Results}
The relation between X-ray photon index, $\Gamma$, and the Eddington-scaled X-ray luminosity, $L_{\rm 2-10 keV}/L_{\rm Edd}$ for all CL AGN sample are shown in Figure 1, where five single sources with multiple observations are plotted in each panel (Mrk 1018, NGC 3516, Mrk 590, NGC 2992 and NGC 1365) and the sources with less observational data points are plotted together in the bottom-right panel. For Mrk 1018, the X-ray spectral index show strong evolution, where $\Gamma$ and $L_{\rm 2-10 keV}/L_{\rm Edd}$ follow positive and negative correlations for $L_{\rm 2-10 keV}/L_{\rm Edd}\textgreater 10^{-3}$ and $\textless 10^{-3}$ respectively. It is interesting to note that the CL AGNs with detected BELs (solid triangles) stay in the positive part while the NEL CL AGNs always exist in the negative part (open triangles). The open circles represent the uncertain type since that we didn't find the quasi-simultaneous optical spectrum observations for given X-ray observations. The top-right panel of Figure 1 shows the results for NGC 3516, where the spectral evolution is not simple as that found in Mrk 1018. The positive correlation between $\Gamma$ and $L_{\rm 2-10 keV}/L_{\rm Edd}$ is evident, while the anti-correlation is not so strong which may be caused by the quite narrow distribution of Eddington ratios. However, BELs always present in the positive part (solid triangles), and the BEL is disappear in low-Eddington-ratio part (open triangles) with a critical Eddington ratio of $\sim 10^{-3}$.  For Mrk 590, 8 data points roughly follow a positive $\Gamma- L_{\rm 2-10 keV}/L_{\rm Edd}$ correlation, where two points with  $L_{\rm 2-10 keV}/L_{\rm Edd}\textless10^{-3}$ follow a roughly negative correlation. It should be noted that NEL and BEL CL AGNs also have the Eddington-scaled X-ray luminosity lower and higher than $10^{-3}$ respectively. It is roughly the similar case for NGC 2992, where the negative correlation is evident for the data points with $L_{\rm 2-10 keV}/L_{\rm Edd}\textless 10^{-3}$. We also present the $\Gamma- L_{\rm 2-10 keV}/L_{\rm Edd}$ relation of NGC 1365, a CL AGN defined by variation of X-ray absorption column density, in bottom-left panel of Figure 1.  It can be found that there are negative and positive correlations between $\Gamma$ and $L_{\rm 2-10 keV}/L_{\rm Edd}$  for $L_{\rm 2-10 keV}/L_{\rm Edd}\textgreater 10^{-3}$ and $\textless 10^{-3}$ respectively, even there are only two observational points in the negative part. For other CL AGNs with quite limited X-ray observations, we present them in bottom-right panel of Figure 1 together, the negative and positive correlations also roughly exist with a critical Eddington ratio of $L_{\rm 2-10 keV}/L_{\rm Edd}\sim 10^{-3}$, where some data observed with BELs also stay in the positive part of  $\Gamma-L_{\rm 2-10 keV}/L_{\rm Edd}$ correlation. For the sample as a whole, we present the $\Gamma-L_{\rm 2-10 keV}/L_{\rm Edd}$ correlation in Figure 2 (the solid squares represent the average value for the given bin). It can found that the negative and positive still exist even for the whole sample, and they transit at $L_{\rm 2-10 keV}/L_{\rm Edd}\sim10^{-3}$.  


In Figure 3, we present the histogram of Eddington-scaled X-ray luminosity for the BEL, NEL and total CL AGNs in our sample. We can find that the BEL and NEL CL AGNs follow a bimodal distribution, where the Eddington ratio of the BEL CL AGNs is evidently higher than that of NEL sources. The whole sample also show the bimodal distribution with the critical Eddington-scaled X-ray luminosity $\sim10^{-3}$.

\section{Conclusion and Discussion}
The CL AGNs challenge our understanding on the AGN unification. In aim to explore the physical mechanism behind the type transition and/or strong variation of column densities, we studied the X-ray properties for a sample of CL AGNs. We find that the X-ray spectral index are negatively and positively correlated with the Eddington-scaled 2-10 keV X-ray luminosity in both a single CL AGN and a sample of CL AGNs when the Eddington ratio is lower and higher than a critical value of $\sim 10^{-3}$, which suggest that the accretion mode is changed during the evolution of CL AGNs.  The CL AGNs with BELs normally stay in the positive part of $\Gamma - L_{\rm 2-10 keV}/L_{\rm Edd}$ correlation, while CL AGNs with only NELs stay in the negative part.  This result suggests that the accretion mode transition should be the physical reason for the changing look.  The histogram distribution of $L_{\rm 2-10 keV}/L_{\rm Edd}$ show a bimodal distribution for BEL and NEL CL AGNs with a critical value of $L_{\rm 2-10 keV}/L_{\rm Edd}\sim 10^{-3}$, which strengthen above conclusions.

The X-ray photon index is strongly correlated to the Eddington ratio in both X-ray binaries (XRBs) and AGNs, where the X-ray photon indices are negatively and positively correlated with the bolometric Eddington ratios when the Eddington ratio is lower and higher than a critical value of $L_{\rm bol}/L_{\rm Edd}\sim 10^{-2}$ respectively  \citep[e.g.,][and references therein]{wa04,ya07,wu08,tr13,yang15}. The strong X-ray luminosity variations are found in many CL AGNs, and it is interesting to note that these CL AGNs follow the similar trends of negative and positive X-ray spectral evolution as that found in both XRBs and other AGN samples.  By assuming the bolometric correction for 2-10 keV X-ray luminosity, $k_{\rm bol}\sim10-30$ \citep[e.g.,][]{marc04}, the critical bolometric Eddington ratio for CL AGNs is roughly similar to that constrained from other AGN sample and XRBs. Our data are selected from different X-ray satellites, and X-ray photon index may be affected a little bit since the different waveband data are adopted ( e.g., typical value of $\Delta\Gamma\sim0.2$), which will not affect our main conclusion. Due to the X-ray luminosity is dominantly come from the accretion flows near the BH horizon, and, therefore, these negative and positive correlations should suggest the possible transition of the accretion mode. The negative correlation is roughly consistent with the prediction from the advection dominated accretion flow (ADAF), where the optical depth of ADAF for Comptonization increases as increasing of accretion rate which will lead to a harder X-ray spectrum \citep[e.g.,][]{gu08,yn14,yang15}. If the accretion rate is larger than a critical value, the cold clumps and/or even cold inner disk will be formed that embedded in the hot accretion flow if optical depth is larger or much larger than one in some places \citep[e.g.,][]{cao09,yo12,qb13,yang15}. If this is the case, it will greatly increase the radiative efficiency of accretion flow and more hot plasma will condense to the cold disk. This process will lead to a softer X-ray spectrum with the decrease of the coronal optical depth.  Our results for the X-ray spectral evolution in the CL AGNs suggest that these sources should be suffered the accretion mode transition. The strong X-ray spectral evolution of CL AGNs is similar to that of XRBs, which is similar to the studies of the correlation between the UV-to-X-ray spectral index \citep[e.g.][]{ruan19,ai19}. The evolution of optical and X-ray observations also suggest the strong evolution of the disk and corona, where the optical emission mainly come from the cold standard disk in bright state.  As shown in Figure 1 \& 2, most BEL CL AGNs stay in the positive correlations while NLE CL AGNs stay in the negative part, which is consistent that as found in other AGNs where the BELs are found in bright Seyferts and QSOs while only weak or no BELs are found in nearby low-luminosity AGNs \citep[e.g.,][]{ho08}.  In low-accretion-rate regime, the standard thin disk may transit to an ADAF at inner region, where the fainter continuum ultimately reduces the emission line intensity of the BLR since there is not enough photons to ionize the gas in the BEL region. Therefore, the BELs will become weaker or even disappear in low-luminosity AGNs. It should be noted that the weak and/or double-peak BELs are found in some low-luminosity AGNs, which suggest that the BELs may become weak but not fully disappear even in low-accretion-rate regime \citep[see][for details]{ho08}. It should be useful to test this issue in CL AGNs with only NELs with higher sensitivity spectral observations. 


The type transition timescale of several CL AGNs is normally several years (e.g., Mrk 1018, NGC 1365 etc.), which is much shorter than the expectation of the flicker between type I and type II due to the orientation effect or the viscous timescale of the accretion disk  \citep[e.g.,][]{nd18}.  To explain this short-time scale in SMBH systems, \cite{sc19} proposed that the radiation-pressure instability operating in a narrow zone between the outer cold disk and an inner ADAF may be responsible for repeating outbursts in some CL AGN, where the outburst timescale is around a couple of years. If the transition radius of SSD and ADAF are small, the X-ray spectrum will also become softer as decrease of transition radius or increase of accretion rate, where the seed photons from outer SSD will become dominant if the transition radius is smaller than several tens gravitational radius. The detailed model calculations are needed to further test this model based on the X-ray spectral evolution. If the accretion rate is close some critical rate, the hot gas will be cooled down in some parts of the ADAF and form the cold clumps and then possibly further form a small cold disk in the inner region of ADAF \citep[e.g.,][]{qb13,yang15}.  

The significant variability of the line-of-sight column density as found in some CL AGNs based on X-ray observations, where such a variability has been discovered on time-scales from a few days down to a few hours (e.g., NGC 1365,  etc.). Such short timescale also suggest that the absorbing material should be at much smaller region compared the conventional torus.  We also explore the X-ray spectral evolution for NGC 1365, which is a CL AGN as defined based on the variation of the column density. It is interesting to note that the X-ray spectral evolution in NGC 1365 is also quite similar to other CL AGNs, where the negative and positive correlations $\Gamma - L_{\rm 2-10 keV}/L_{\rm Edd}$ are found when the Eddington-scale X-ray luminosity is less and larger than  $\sim 10^{-3}$ respectively. This result suggest that the accretion mode also changed in this X-ray defined CL AGN, and the variable absorbing material should be correlated with the evolution of the accretion mode.  As discussed above, the clumpy cold gas or even cold disk may be formed when accretion rate is larger than a critical value. This clumpy cold gas can be serve as the cold absorbing gas if the inclination angle is close to plane of the cold clumpy gas.

\section*{Acknowledgements}
This work is supported by the NSFC (grants U1931203,11622324, and 11573009) . We appreciate for the useful discussions with Wu Xuebing, Yu Wenfei, Cao Xinwu, and Gu Minfeng. 

\facility{NuSTAR, XMM-Newton, Chandra}
\software{HEASOFT (v6.25), XSPEC (v12.10.1), SAS (v18.0.0), CIAO (v4.11)}

\newpage

\begin{center}
\centering

\centerline{Table 1: the basic parameters of CL-AGN sample.}
\footnotesize
\begin{longtable}{llcccc}
  \hline
  \hline
  Source & Redshift & log(M$_{\rm BH}$/$\msun$) & Ref. & Defining Method & Number     \\
  ~[1]&[2]&[3]&[4]&[5]&[6]   \\
  \hline
  \hline
\endhead
  Mrk 1018 & 0.0424 & 8.25 & a & A & 23   \\
  NGC 3516 & 0.0088 & 7.49 & b & A & 17   \\
  Mrk 590 & 0.0264 & 7.50 & c & A & 10   \\
  NGC 2992 & 0.0077 & 7.72 & d & A & 14   \\
  NGC 1365 & 0.0055 & 6.65 & e & B & 13   \\
  NGC 1566 & 0.0050 & 6.92 & d & A & 2   \\
  HE 1136-2304 & 0.0270 & 7.58 & f & A & 1   \\
  NGC 7582 & 0.0053 & 7.25 & g & A & 5   \\
  SDSS J015957.64+003310.4 & 0.3120 & 8.15 & h & A & 1   \\
  NGC 4151 & 0.0033 & 7.56 & b & A & 13   \\
  Mrk 530(NGC 7603) & 0.0295 & 8.06 & i & A & 2   \\
  Mrk 273 & 0.0378 & 7.74 & j & B & 4   \\
  NGC 7319 & 0.0225 & 7.38 & d & B & 3   \\
  IC 751 & 0.0311 & 8.50 & k & B & 1   \\
  UGC 4203(Mrk 1210) & 0.0135 & 6.78 & l & B & 3   \\
  \hline
  \hline
\end{longtable}
\begin{minipage}{170mm}
Column [1]: Source names; Columns [2]: Redshift; Column [3] and [4]: The BH mass and reference; Column [5]: A and B represent the CL AGNs defined from the change of BEL/NEL and the change of X-ray column density respectively; Column [6]: The number of observations.   \\
The references of SMBH masses in Column [4]: a. \cite{wint10}; b. \cite{grie13}; c. Calculated from the scaling relations in \cite{mcco13} and central velocity dispersion of source from the web: http://leda.univ-lyon1.fr/; d. \cite{woo02}; e. \cite{faus18}; f. \cite{koll18}; g. \cite{lama10}; h. \cite{ruan19}; i. \cite{ehle18}; j. \cite{gonz09}; k. \cite{ricc16}; l. \cite{onor17}. \\

\end{minipage}
\label{para}
\end{center}

\newpage

\begin{center}
\centering

\centerline{Table 2: the fitting results of CL-AGN samples.}
\footnotesize
\begin{longtable}{lcllclccl}
  \hline
  \hline
  Year & Telescope & Model & N$_{\rm H}$ & $\Gamma$  &  log($L_{\rm 2-10}$  &  $\chi^{2}$/dof.  &  Type\ddag & Obsid/Ref.   \\
   &  &  &  ($10^{22}$) &  &  /$L_{\rm Edd}$)  &  &  &   \\
  ~[1]&[2] &[3] &[4] &[5] &[6] &[7] &[8] &[9]  \\
  \hline
  \hline
\endhead
  \multicolumn{9}{c}{\textbf{Mrk 1018}\dag}     \\
  2005 & Sw & A & 0.02 & $1.93 _{ -0.05 }^{+ 0.05 }$  &  -2.68 & 1.10 & BEL & (a)   \\
  2005 & X & B & 0.02 & $1.83 _{ -0.05 }^{+ 0.04 }$ &  -2.79 & 0.90 & BEL & 0201090201   \\
  2007 & Sw & A & 0.02 & $1.91 _{ -0.08 }^{+ 0.08 }$  &  -2.77 & 1.10 & U & (a)   \\
  2007 & Sw & A & 0.02 & $1.95 _{ -0.08 }^{+ 0.08 }$  &  -2.84 & 1.20 & U & (a)   \\
  2007 & Sw & A & 0.02 & $1.95 _{ -0.07 }^{+ 0.07 }$  &  -2.80 & 1.00 & U & (a)   \\
  2008 & Sw & A & 0.02 & $1.76 _{ -0.06 }^{+ 0.06 }$  &  -2.74 & 1.00 & BEL & (a)   \\
  2008 & X & B & 0.02 & $1.95 _{ -0.02 }^{+ 0.02 }$ &  -2.68 & 1.14 & BEL & 0554920301   \\
  2009 & Su & C & 0.02 & $2.00 _{ -0.03 }^{+ 0.03 }$  &  -2.73 & 1.19 & BEL & (b)   \\
  2010 & C & B & 0.02 & $1.67 _{ -0.03 }^{+ 0.03 }$ &  -2.76 & 1.21 & BEL & 12868   \\
  2013 & Sw & A & 0.02 & $1.42 _{ -0.18 }^{+ 0.18 }$  &  -2.83 & 1.10 & U & (a)   \\
  2014 & Sw & A & 0.02 & $1.50 _{ -0.60 }^{+ 0.60 }$  &  -3.53 & 0.50 & NEL & (a)   \\
  2016 & Sw & A & 0.02 & $1.75 _{ -0.27 }^{+ 0.27 }$  &  -3.55 & 1.30 & NEL & (a)   \\
  2016 & Sw & A & 0.02 & $1.33 _{ -0.26 }^{+ 0.26 }$  &  -3.33 & 0.50 & NEL & (a)   \\
  2016 & C & B & 0.02 & $1.72 _{ -0.05 }^{+ 0.05 }$ &  -3.66 & 0.90 & NEL & 18789   \\
  2016 & N & B & 0.02 & $1.84 _{ -0.16 }^{+ 0.17 }$ &  -3.53 & 0.89 & NEL & 60160087002   \\
  2017 & C & B & 0.02 & $1.61 _{ -0.03 }^{+ 0.03 }$ & -3.33 & 1.25 & NEL & 19560   \\
  2018 & C & B & 0.02 & $1.60 _{ -0.06 }^{+ 0.05 }$ &  -3.49 & 1.02 & U & 20366   \\
  2018 & N & B & 0.02 & $1.79 _{ -0.11 }^{+ 0.12 }$ &  -3.48 & 1.02 & U & 60301022002   \\
  2018 & N & B & 0.02 & $1.72 _{ -0.10 }^{+ 0.11 }$ &  -3.57 & 1.09 & U & 60301022003   \\
  2018 & N & B & 0.02 & $1.68 _{ -0.09 }^{+ 0.09} $ &  -3.46 & 1.06 & U & 60301022005   \\
  2018 & C & B & 0.02 & $1.63 _{ -0.06 }^{+ 0.06 }$ &  -3.56 & 0.98 & U & 20367   \\
  2018 & C & B & 0.02 & $1.60 _{ -0.06 }^{+ 0.05 }$ &  -3.49 & 1.01 & U & 20368   \\
  2018 & C & B & 0.02 & $1.61 _{ -0.05 }^{+ 0.04 }$ &  -3.31 & 1.07 & U & 20369   \\
  \cline{1-9} 
  \multicolumn{9}{c}{\textbf{NGC 3516}}     \\
  2001 & X & D & $1.39_{-0.09}^{+0.10}$ & $1.58 _{ -0.03 }^{+ 0.03 }$ &  -2.95 & 1.10 & BEL & 0107460601   \\
  2001 & X & D & $1.49_{-0.09}^{+0.11}$ & $1.32 _{ -0.03 }^{+ 0.03 }$ &  -3.11 & 1.19 & BEL & 0107460701   \\
  2005 & Su & E & $5.50_{-0.20}^{+0.20}$ & $1.90 _{ -0.03 }^{+ 0.03 }$  &  -2.86 & 1.05 & BEL & (c)   \\
  2006 & X & D & $0.82_{-0.07}^{+0.06}$ & $1.85 _{ -0.02 }^{+ 0.02 }$ &  -2.58 & 1.17 & BEL & 0401210401   \\
  2006 & X & D & $0.86_{-0.07}^{+0.07}$ & $1.84 _{ -0.02 }^{+ 0.02 }$ &  -2.64 & 1.17 & BEL & 0401210501   \\
  2006 & X & D & $1.40_{-0.09}^{+0.08}$ & $1.77 _{ -0.03 }^{+ 0.02 }$ &  -2.72 & 1.21 & BEL & 0401210601   \\
  2006 & X & D & $0.86_{-0.07}^{+0.08}$ & $1.86 _{ -0.02 }^{+ 0.03 }$ &  -2.64 & 1.10 & BEL & 0401211001   \\
  2009 & Su & F & $3.21_{-1.65}^{+0.55}$ & $1.73 _{ -0.06 }^{+ 0.03 }$  &  -3.21 & 1.08 & U & (d)   \\
  2013 & Su & G & 0.88-2.31 & $1.75 _{ -0.02 }^{+ 0.01 }$  &  -4.01 & 1.06 & NEL & (e)   \\
  2014 & N & G & $4.20_{-0.59}^{+2.49}$ & $1.77 _{ -0.05 }^{+ 0.12 }$  & -3.62 & 1.05 & NEL & 60002042002   \\
  2014 & N & G & $2.20_{-0.45}^{+0.46}$ & $1.71 _{ -0.04 }^{+ 0.07 }$ &  -3.50 & 1.19 & NEL & 60002042004   \\
  2017 & N & G & $1.45_{-0.73}^{+0.78}$ & $1.61 _{ -0.06 }^{+ 0.12 }$ &  -3.57 & 0.87 & NEL & 60302016002   \\
  2017 & N & G & $3.85_{-1.05}^{+1.11}$ & $1.77 _{ -0.06 }^{+ 0.12 }$ &  -3.55 & 0.83 & NEL & 60302016004   \\
  2017 & N & G & $1.70_{-0.96}^{+1.02}$ & $1.64 _{ -0.06 }^{+ 0.11 }$ &  -3.56 & 1.22 & NEL & 60302016006   \\
  2017 & N & G & $1.23_{-0.81}^{+0.84}$ & $1.78 _{ -0.10 }^{+ 0.07 }$ &  -3.30 & 0.98 & NEL & 60302016008   \\
  2017 & N & G & $<0.01$ & $1.62 _{ -0.06 }^{+ 0.08 }$ &  -3.33 & 0.97 & NEL & 60302016010   \\
  2017 & N & G & $1.88_{-0.97}^{+1.00}$ & $1.77 _{ -0.08 }^{+ 0.07 }$ &  -3.18 & 0.88 & NEL & 60302016012   \\
  \cline{1-9} 
  \multicolumn{9}{c}{\textbf{Mrk 590}\dag}     \\
  2000 & R & H & 0.03 & $1.75 _{ -0.08 }^{+ 0.08 }$  &  -2.41 & 0.63 & U & (f)   \\
  2002 & X & B & 0.03 & $1.73 _{ -0.03 }^{+ 0.04 }$ &  -2.74 & 0.98 & BEL & 0109130301   \\
  2004 & C & B & 0.03 & $1.63 _{ -0.05 }^{+ 0.05 }$ &  -2.63 & 1.05 & BEL & 4924   \\
  2004 & X & D & 0.03 & $1.75 _{ -0.02 }^{+ 0.01 }$ &  -2.56 & 1.07 & BEL & 0201020201   \\
  2011 & Su & I & 0.03 & $1.67 _{ -0.01 }^{+ 0.01 }$  &  -2.55 & 0.91 & U & (g)   \\
  2013 & C & B & 0.03 & $1.79 _{ -0.13 }^{+ 0.13 }$ &  -3.52 & 0.93 & NEL & 15647   \\
  2014 & C & B & 0.03 & $1.68 _{ -0.09 }^{+ 0.08 }$ &  -3.37 & 1.14 & NEL & 16109   \\
  2016 & N & B & 0.03 & $1.66 _{ -0.09 }^{+ 0.09 }$ &  -2.88 & 0.90 & U & 60160095002   \\
  2016 & N & B & 0.03 & $1.61 _{ -0.06 }^{+ 0.07 }$ &  -2.93 & 0.97 & U & 90201043002   \\
  2018 & N & B & 0.03 & $1.68 _{ -0.04 }^{+ 0.05 }$ &  -2.38 & 0.86 & U & 80402610002   \\
  \cline{1-9} 
  \multicolumn{9}{c}{\textbf{NGC 2992}\dag}     \\
  2003 & X & D & $0.68_{-0.02}^{+0.03}$ & $1.80 _{ -0.04 }^{+ 0.18 }$ &  -2.67 & 1.00 & U & 0147920301   \\
  2005 & Su & J & $0.80_{-0.05}^{+0.06}$ & $1.57 _{ -0.03 }^{+ 0.06 }$  &  -3.63 & 1.08 & NEL & (h)   \\
  2005 & I & K & - & $1.96 _{ -0.23 }^{+ 0.26 }$  &  -2.89 & - & BEL & (i)   \\
  2005 & R & H & - & $1.78 _{ -0.18 }^{+ 0.18 }$  &  -3.86 & 0.44 & NEL & (f)   \\
  2010 & X & D & 0.85 & $1.59 _{ -0.02 }^{+ 0.02 }$ &  -3.87 & 1.10 & U & 0654910301   \\
  2010 & X & D & 0.85 & $1.62 _{ -0.01 }^{+ 0.02 }$ &  -3.82 & 1.07 & U & 0654910401   \\
  2010 & X & D & 0.85 & $1.59 _{ -0.01 }^{+ 0.02 }$ &  -3.54 & 1.03 & U & 0654910501   \\
  2010 & X & D & 0.85 & $1.63 _{ -0.02 }^{+ 0.03 }$ &  -3.96 & 1.21 & U & 0654910601   \\
  2010 & X & D & 0.85 & $1.65 _{ -0.02 }^{+ 0.02 }$ &  -3.95 & 1.26 & U & 0654910701   \\
  2010 & X & D & 0.85 & $1.65 _{ -0.03 }^{+ 0.03 }$ &  -4.05 & 1.24 & U & 0654910801   \\
  2010 & X & D & 0.85 & $1.64 _{ -0.03 }^{+ 0.04 }$ &  -4.26 & 1.48 & U & 0654910901   \\
  2010 & X & D & 0.85 & $1.64 _{ -0.02 }^{+ 0.01 }$ &  -3.92 & 1.08 & U & 0654911001   \\
  2013 & X & D & 0.85 & $1.59 _{ -0.02 }^{+ 0.02 }$ &  -3.46 & 1.02 & U & 0701780101   \\
  2015 & N & D & 0.85 & $1.73 _{ -0.01 }^{+ 0.02 }$ &  -2.88 & 0.95 & U & 60160371002   \\
  \cline{1-9} 
  \multicolumn{9}{c}{\textbf{NGC 1365}}     \\
  2002 & C & L & - & $1.98 _{ -0.14 }^{+ 0.13 }$  &  -3.61 & 1.09 & U & (j)   \\
  2003 & X & M & $47.60_{-0.50}^{+3.00}$ & $2.06 _{ -0.03 }^{+ 0.06 }$  &  -2.67 & 1.09 & U & (j)   \\
  2006 & X & L & - & $2.09 _{ -0.14 }^{+ 0.11 }$  &  -3.72 & 1.06 & U & (j)   \\
  2006 & X & M & $33.50_{-2.00}^{+1.90}$ & $2.33 _{ -0.12 }^{+ 0.12 }$  &  -2.40 & 1.21 & U & (j)   \\
  2006 & C & N & $40.00_{-4.00}^{+5.00}$ & $2.10 _{ -0.25 }^{+ 0.25 }$  &  -2.21 & 1.02 & U & (k)   \\
  2008 & Su & O & $15.40_{-0.20}^{+0.30}$ & $1.81 _{ -0.04 }^{+ 0.04 }$  &  -2.56 & 1.15 & U & (l)   \\
  2010 & Su & O & $58.30_{-0.60}^{+0.50}$ & $1.76 _{ -0.03 }^{+ 0.02 }$  &  -2.58 & 1.15 & U & (l)   \\
  2010 & Su & O & $106.60_{-3.20}^{+5.30}$ & $1.73 _{ -0.13 }^{+ 0.01 }$  &  -2.71 & 1.15 & U & (l)   \\
  2012 & N & P & $13.24_{-1.36}^{+1.43}$ & $1.85 _{ -0.12 }^{+ 0.12 }$ & -2.79 & 1.05 & U & 60002046002   \\
  2012 & N & P & $14.04_{-1.41}^{+1.46}$ & $1.74 _{ -0.13 }^{+ 0.12 }$ & -2.72 & 0.98 & U & 60002046003   \\
  2012 & N & P & $2.30_{-0.60}^{+0.59}$ & $1.87 _{ -0.08 }^{+ 0.07 }$ &  -2.47 & 1.38 & U & 60002046005   \\
  2013 & N & P & $<0.01$ & $1.94 _{ -0.06 }^{+ 0.07 }$ &  -2.48 & 1.28 & U & 60002046007   \\
  2013 & N & P & $7.77_{-0.69}^{+0.69}$ & $1.99 _{ -0.08 }^{+ 0.07 }$ &  -2.53 & 1.02 & U & 60002046009   \\
  \multicolumn{9}{c}{\textbf{Others}}     \\
  \multicolumn{9}{c}{NGC 1566}     \\
  2012 & Su/Sw & Q & $3.04_{-0.82}^{+1.08}$ & $2.04 _{ -0.09 }^{+ 0.10 }$  &  -4.07 & 1.18 & NEL & (m)   \\
  2018 & N & R & $1.00_{-0.36}^{+0.36}$ & $1.94 _{ -0.05 }^{+ 0.04 }$ &  -2.44 & 0.95 & BEL & 80301601002   \\
  \multicolumn{9}{c}{HE 1136-2304}     \\
  2014 & N & B & $1.48_{-0.92}^{+0.95}$ & $1.75 _{ -0.05 }^{+ 0.04 }$ &  -2.47 & 1.06 & BEL & 80002031003   \\
  \multicolumn{9}{c}{NGC 7582}     \\
  2003 & R & S & unconstr. & $2.10 _{ -0.10 }^{+ 0.10 }$  &  -3.54 & 0.87 & U & (n)   \\
  2007 & Su & T & $120.00_{-20.00}^{+20.00}$ & $1.92 _{ -0.16 }^{+ 0.24 }$  &  -2.90 & 0.97 & U & (o)   \\
  2012 & N & B & $21.45_{-4.10}^{+4.68}$ & $1.21_{-0.09}^{+0.10}$ &  -3.54 & 1.05 & U & 60061318002   \\
  2012 & N & B & $39.38_{-7.71}^{+8.78}$ & $1.24 _{ -0.13 }^{+ 0.13 }$ &  -3.55 & 1.08 & U & 60061318004   \\
  2016 & N & B & $21.43_{-1.42}^{+1.38}$ & $1.38 _{ -0.03 }^{+ 0.04 }$ &  -3.21 & 1.12 & U & 60201003002   \\
  \multicolumn{9}{c}{SDSS J015957.64+003310.4}     \\
  2000 & X & U & $<0.01$ & $2.22 _{ -0.08 }^{+ 0.09 }$  &  -2.48 & 1.40 & BEL & (p)   \\
  \multicolumn{9}{c}{NGC 4151}     \\
  2000 & X & D & $4.52_{-0.17}^{+0.18}$ & $1.22 _{ -0.04 }^{+ 0.03 }$ &  -3.51 & 1.02 & U & 0112310101   \\
  2003 & X & D & $4.50_{-0.11}^{+0.11}$ & $1.42 _{ -0.02 }^{+ 0.03 }$ &  -2.82 & 1.13 & U & 0143500101   \\
  2003 & X & D & $4.40_{-0.10}^{+0.09}$ & $1.44 _{ -0.02 }^{+ 0.02 }$ &  -2.82 & 1.10 & U & 0143500201   \\
  2003 & X & D & $3.93_{-0.08}^{+0.08}$ & $1.48 _{ -0.02 }^{+ 0.02 }$ &  -2.73 & 1.16 & U & 0143500301   \\
  2006 & X & D & $6.34_{-0.15}^{+0.15}$ & $1.29 _{ -0.02 }^{+ 0.03 }$ &  -3.19 & 1.10 & U & 0402660201   \\
  2011 & X & D & $4.93_{-0.17}^{+0.18}$ & $1.50 _{ -0.03 }^{+ 0.03 }$ &  -2.87 & 1.03 & U & 0657840301   \\
  2011 & X & D & $4.02_{-0.16}^{+0.16}$ & $1.34 _{ -0.04 }^{+ 0.03 }$ &  -2.90 & 1.08 & U & 0657840401   \\
  2012 & X & D & $8.05_{-0.24}^{+0.26}$ & $1.28 _{ -0.04 }^{+ 0.05 }$ &  -2.81 & 1.20 & U & 0679780101   \\
  2012 & X & D & $5.70_{-0.11}^{+0.10}$ & $1.29 _{ -0.06 }^{+ 0.05 }$ &  -3.01 & 1.14 & U & 0679780301   \\
  2012 & X & D & $6.03_{-0.17}^{+0.20}$ & $1.37 _{ -0.04 }^{+ 0.03 }$ &  -2.88 & 1.07 & U & 0679780401   \\
  2012 & N & V & $5.25_{-0.65}^{+0.63}$ & $1.44 _{ -0.04 }^{+ 0.04 }$ &  -2.87 & 1.17 & U & 60001111002   \\
  2012 & N & V & $7.28_{-0.45}^{+0.46}$ & $1.53 _{ -0.07 }^{+ 0.07 }$ &  -2.89 & 1.06 & U & 60001111003   \\
  2012 & N & V & $5.65_{-0.44}^{+0.40}$ & $1.46 _{ -0.04 }^{+ 0.05 }$ &  -2.91 & 1.26 & U & 60001111005   \\
  \multicolumn{9}{c}{Mrk 530(NGC 7603)}     \\
  2012 & Su & W & 0.04\dag & $1.97 _{ -0.02 }^{+ 0.02 }$  &  -2.56 & 1.03 & U & (q)   \\
  2012 & Su & X & $22.00_{-3.00}^{+7.00}$ & $2.22 _{ -0.06 }^{+ 0.06 }$  &  -2.20 & 0.88 & U & (q)   \\
  \multicolumn{9}{c}{Mrk 273}     \\
  2000 & C/X/Su & Y & $159.10_{-7.30}^{+8.60}$ & $1.73 _{ -0.02 }^{+ 0.04 }$  &  -2.40 & 1.39 & U & (r)   \\
  2006 & Su & Z & $8.64_{-0.80}^{+3.71}$ & $1.56 _{ -0.09 }^{+ 0.15 }$  &  -3.01 & 1.25 & U & (r)   \\
  2013 & N & B & $21.49_{-6.89}^{+8.57}$ & $1.22 _{ -0.15 }^{+ 0.16 }$ &  -3.11 & 0.95 & U & 60002028002   \\
  2016 & C & D & $<0.01$ & $1.56 _{ -0.22 }^{+ 0.27 }$ &  -3.93 & 1.16 & U & 18177   \\
  \multicolumn{9}{c}{NGC 7319}     \\
  2000 & C & AA & $39.11_{-5.46}^{+9.32}$ & $1.29 _{ -0.50 }^{+ 1.09 }$ &  -2.67 & 1.53 & U & (s)   \\
  2001 & X & AA & $51.88_{-5.68}^{+5.81}$ & $1.35 _{ -0.28 }^{+ 0.28 }$ &  -2.91 & 1.24 & U & (s)   \\
  2007 & C & AB & $46.06_{-2.91}^{+2.95}$ & $2.03 _{ -0.24 }^{+ 0.22 }$ &  -2.43 & 1.33 & U & (s)   \\
  \multicolumn{9}{c}{IC 751}     \\
  2014 & N & B & $17.40_{-5.47}^{+6.64}$ & $1.48 _{ -0.18 }^{+ 0.20 }$ &  -3.84 & 0.89 & U & 60001148002   \\
  \multicolumn{9}{c}{UGC 4203(Mrk 1210)}    \\
  2001 & X & AC & $17.60_{-2.10}^{+1.30}$ & $1.62 _{ -0.23 }^{+ 0.11 }$ &  -1.84 & 1.21 & U & (t)   \\
  2007 & Su & AC & $33.00_{-2.30}^{+2.20}$ & $1.87 _{ -0.18 }^{+ 0.18 }$ &  -1.97 & 1.14 & U & (t)   \\
  2012 & N & R & $20.85_{-3.54}^{+3.14}$ & $1.95 _{ -0.32 }^{+ 0.09 }$ & -1.97 & 1.01 & U & 60061078002   \\

  \hline
  \hline
\end{longtable}
\begin{minipage}{170mm}
Column [1] and [2]: The observation year and telescopes (C for Chandra; I for INTEGRAL; N for NuSTAR; R for RXTE; Su for Suzaku; Sw for Swift; and X for XMM-Newton ); Column [3]: Fitting models; Column [4]: The column density of neutral hydrogen gas with the unit as $cm^{-2}$; Column [5]: photon index ($\Gamma$); Column [6]: 2-10 keV X-ray scaled Eddington ratios; Column [7]: The reduced $\chi^{2}$ value; Column [8]: The types of optical spectrum (BEL for broad emission line; NEL for only narrow emission lines; U for uncertain for given X-ray observation); Column [9]: The observation ID or the reference if select from literatures.    \\
The fitting models in Column [3]: A: a power-law puls Galactic absorption component; B: pha(pow); C: tbabs(zbbody+pexrav+gau); D: pha(pow+zgau); E: 1-absorber + Compton reflection + diskline + high-ionization absorber; F: absorption(wabs) with power-law, reflection power-law (pexrav) and gaussian line; G: pha(cutoff+pexmon); H: pha(pow+gau+pexrav); I: a power law with Galactic absorption,  an Fe K$\alpha$ line and Compton reflection; J: model with apec, diskline, hrefl and gau; K: a single power law; L: pexrav+gau; M: pexrav+gau with an absorbed power law; N: vapec(zpow+pexrav+zgau*3); O: TBabs*(2vapec+4zgauss+WA*(reflionx+zpcfabs*(powerlaw+zgauss+relconv(reflionx)))); P: pha(refl*pow+zgau); Q: constant*zpcfabs*(zpowerlw*zhighect+rdblur*pexmon)+pexmon+apec; R: pha(pexrav); S: absorption with pexrav and gaussian line; T: highly obscured powerlaw, pexrav, a Gaussian line to model iron K$\alpha$ emission; U: wabs(pow); V: pha(pexrav+zgau); W: const*tbabs*(cutoffpl+const*relxill+pexmon+zgauss+zgauss); X: const*tbabs*(zxipcf*zxipcf*cutoffpl+pexmon+zgauss+zgauss); Y: an absorbed power law , pow, MEKAL and covering fraction; Z: an absorbed power law, pow and MEKAL; AA: pha(zpha*pow+mekal+zpha*pow); AB: pha(zpha*pow+mekal+mekal+zpha*pow); AC: an absorber at the redshift, a power law, a Compton reflection component, a neutral and narrow iron line.   \\
The references in Column [9]: (a) \cite{huse16}; (b) \cite{wint12}; (c) \cite{mark08}; (d) \cite{patr11}; (e) \cite{noda16}; (f) \cite{rive13}; (g) \cite{rive12}; (h) \cite{yaqo07}; (i) \cite{beck07}; (j) \cite{risa06}; (k) \cite{vent18}; (l) \cite{bren13}; (m) \cite{kawa13}; (n) \cite{rive15}; (o) \cite{bian10}; (p) \cite{corr11}; (q) \cite{ehle18}; (r) \cite{teng09}; (s) \cite{hern15}; (t) \cite{matt09}.       \\
 \dag : The column density of neutral hydrogen gas is fixed, where the hydrogen column density of Mrk 1018, Mrk 590 and NGC 2992 is selected from \citep[see ][]{kalb05}  and  \citep[e.g., ][]{mari18} respectively. \\
\ddag : The AGN spectrum types are defined by the optical spectra in following references as listed below: Mrk 1018 (\cite{mcel16}); NGC 3516 (\cite{shap19}); Mrk 590 (\cite{denn14}); NGC 2992 (\cite{trip08}); NGC 1566 (\cite{park19}); HE 1136-2304 (\cite{park16}); SDSS J015957.64+003310.4 (\cite{ruan19}).   \\

\end{minipage}
\label{para}
\end{center}

\newpage

\begin{figure}
  \centering
  \includegraphics[height=0.6\textheight]{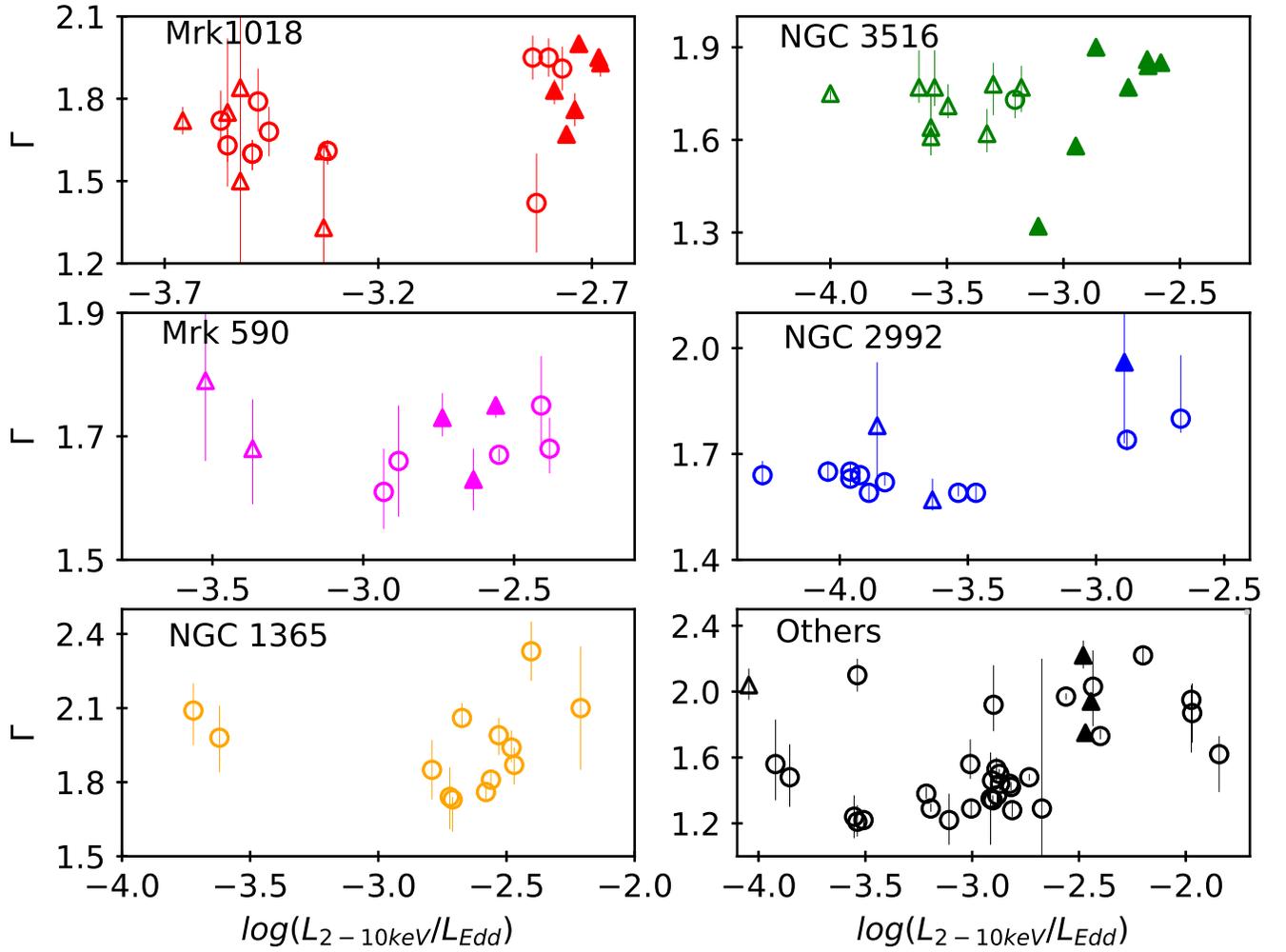} \\
  \caption{The $\Gamma-L_{\rm 2-10 keV}/L_{\rm Edd}$ relation for Mrk 1018, NGC 3516, Mrk 590, NGC 2992, NGC 1365 and other AGN sample respectively. The solid triangles, open triangles and open circles represent the CL AGNs observed with broad emission lines (BELs), without BELs and uncertain type due to no quasi-simultaneous optical spectrum data respectively.}
\end{figure}

\begin{figure}
  \centering
  \includegraphics[height=0.4\textheight]{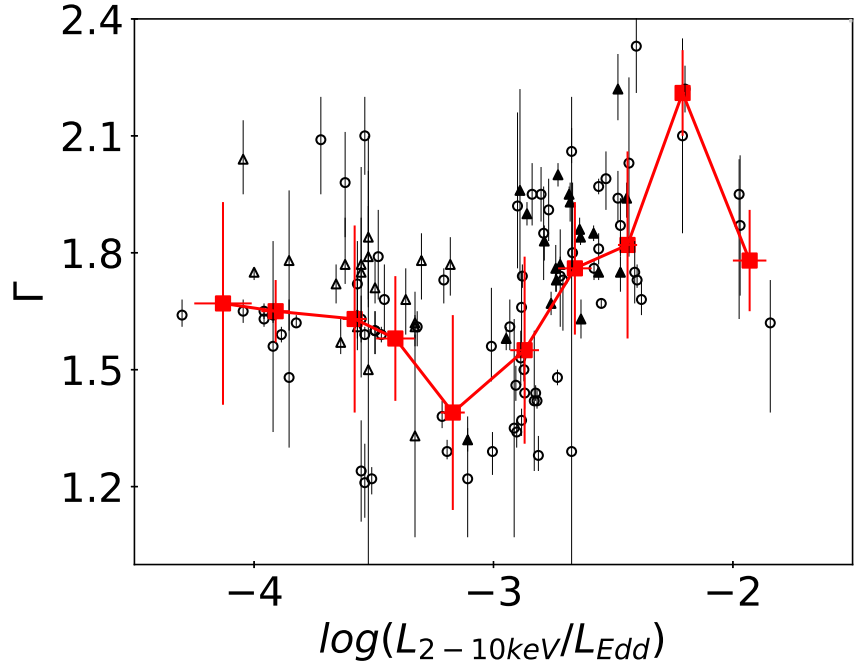} \\
  \caption{The $\Gamma-L_{\rm 2-10 keV}/L_{\rm Edd}$ relation for all CL AGN sample. The symbols are the same as Figure 1 except that the solid squares linked with solid lines represent the average value at given bin.}

\end{figure}

\begin{figure}
  \centering
  \includegraphics[height=0.4\textheight]{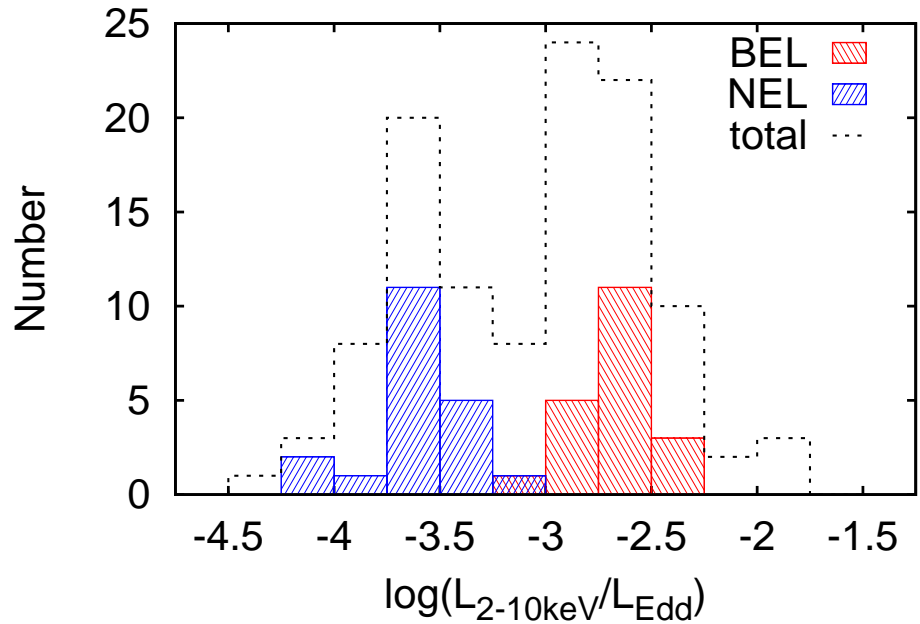} \\
  \caption{The histogram of the X-ray scaled Eddington ratio for all CL AGNs in this work.}

\end{figure}

\end{document}